\documentclass[pra,twocolumn,aps]{revtex4}
\usepackage{graphicx}
\usepackage{graphicx}
\usepackage{epsfig}
\usepackage{dcolumn}
\usepackage{bm}
\usepackage{bbm}
\usepackage{amsmath}
\usepackage{graphicx}
\usepackage{amsfonts}
\usepackage{hyperref}
\usepackage{amssymb}
\usepackage{color}
\usepackage[up]{subfigure}
\newtheorem{thm}{Theorem}

\newcommand{\be}{\begin{equation}}
\newcommand{\ee}{\end{equation}}
\newcommand{\bc}{\begin{center}}
\newcommand{\ec}{\end{center}}
\newcommand{\bea}{\begin{eqnarray}}
\newcommand{\eea}{\end{eqnarray}}

\newcommand{\halfsq}{\frac{1}{\sqrt{2}}}

\begin{document}
\title{Quantifying  quantumness  via  commutators: an  application  to
  quantum   walks}   \author{Pavan   Iyengar}   
\email{pavan.iyengar@students.iiserpune.ac.in}  \affiliation{  NMR-QIP
  Lab,  Physics  Deptartment,  IISER,  Pune, India}  \author{G.   N.
  Chandan}    \email{chandan.linus@gmail.com}    \affiliation{   Raman
  Research   Institute,   Bengaluru,   India}  \author{R.    Srikanth}
\email{srik@poornaprajna.org} \affiliation{  Poornaprajna Institute of
  Scientific Research, Bengaluru, India}


\begin{abstract}
The question  of witnessing or quantifying  nonclassicality of quantum
systems has  been addressed  in various ways.   For a given  system or
theory,  we  propose  identifying   it  with  the  incompatibility  of
admissible states.  We quantify the nonclassicality of quantum systems
using the Hilbert-Schmidt norm of  the commutator of two states.  As a
particular application of this  measure, we study the classicalization
of a discrete-time quantum walk with a noisy coin.
\end{abstract}

\maketitle

\section{Introduction}

There are a number of  ways to characterize the nonclassical nature of
quantum  phenomena.   In   connection  with  the  quantum  measurement
problem, the lack of  macroscopic superpositions is the tell-tale sign
of classicality \cite{Zeh03,  Sch05}. Put differently, the macro-world
is classical because the accessible states all commute with each other
(being positional eigenstates). We can extend this idea farther to any
system $S$  (and even to  a theory $T$),  defining it to  be classical
precisely  if  all  admissible  states  of $S$  or  $T$  are  mutually
compatible.

What  is advantageous  in  this  approach is  that  references to  the
dynamics  and  correlations are  removed,  which  can offer  potential
simplification  and  straightforwardness  for quantification.   As  an
illustration:  maximal  entanglement,   which  is  often  regarded  as
quintessential nonclassicality, may be  represented as a product state
and vice versa, simply by a  suitable choice of the degrees of freedom
(cf.  Ref.  \cite{ZLL04}).  For a  2-qubit system, suppose we define a
parity    observable   by    the   spectral    decomposition    $P   =
\left(\Pi_{\Phi^+}   +  \Pi_{\Phi^-}\right)  -   \left(\Pi_{\Psi^+}  +
\Pi_{\Psi^-}\right)$    and   the   phase    observable   by    $F   =
\left(\Pi_{\Phi^+}   +  \Pi_{\Psi^+}\right)  -   \left(\Pi_{\Phi^-}  +
\Pi_{\Psi^-}\right)$, where  $\Pi_{\Phi^\pm}$ is the  projector to the
Bell state  $\halfsq(|00\rangle \pm |11\rangle)$,  and $\Pi_{\Psi^\pm}
\equiv \halfsq(|01\rangle  \pm |10\rangle)$.   In terms of  the parity
and  phase  observables,  the  Bell  states  have  the  product  form:
$|0\rangle_P|0\rangle_F  = |\Phi^+\rangle$,  $|0\rangle_P|1\rangle_F =
|\Phi^-\rangle$,   $|1\rangle_P|0\rangle_F   =  |\Psi^+\rangle$,   and
$|1\rangle_P|1\rangle_F  = |\Psi^-\rangle$.   On the  other  hand, the
separable  state  $\frac{1}{\sqrt{2}}(|0\rangle +  |1\rangle)|0\rangle
\equiv          \frac{1}{2}\left[|0\rangle_P\left(|0\rangle          +
  |1\rangle\right)_F     +     |1\rangle_P\left(|0\rangle_\varphi    -
  |1\rangle\right)_F\right]          \equiv         \frac{1}{\sqrt{2}}
\left(|0\rangle_P|{+}\rangle_F + |1\rangle_P|{-}\rangle_F\right)$.

As  one way to  overcome this  problem, which  may be  appropriate for
systems   of   the   type   considered  below,   we   propose   basing
nonclassicality  on the  incompatibility  of states  relative to  each
other, rather than on correlations.  We believe that this approach can
be helpful in  studying the quantumness of complex  processes, such as
those  encountered   in  photosynthetic  systems,  where   it  may  be
computationally  unfeasible  to  compute measures  of  nonclassicality
based on  correlations.  That  the experiments in  photosynthesis show
quantum  effects \cite{NBT11},  while rivaling  classical explanations
for exciton transport also exist, suggests that our approach will find
fruitful  application  here.   In  particular, evidence  of  long-term
coherence \cite{Eng+07} and of continuous quantum walks \cite{MRLG08},
have been  reported to play a  role in the energy  transfer within the
Fenna-Mathews-Olson  (FMO) complex.  Accordingly  we have  applied our
method to quantum walks.

This article  is organized as  follows.  In Section  \ref{sec:def}, we
develop the  conceptual background for  identifying nonclassicality in
terms of  the incompatibility of  states. In Section  \ref{sec:HS}, we
propose a  method for  quantifying nonclassicality in  quantum systems
using the commutator as the  basic witness of quantumness.  In Section
\ref{sec:QW}, we  apply this approach to  discrete-time quantum walks.
We conclude in Section \ref{sec:C}.

\section{Nonclassicality from Incompatibility of states \label{sec:def}}

Given a system  $S$, let $\Sigma \equiv \{\psi_j\}$ be  the set of all
possible  pure states  that  it can  assume  in a  given situation  of
interest. Any two pure states  $\psi_m$ and $\psi_n$ (where $m, n$ are
indices appropriate  to the  cardinality of $\Sigma$)  are said  to be
incompatible    if     their    \textit{characteristic    properties},
$\Pi(\psi_m)$ and $\Pi(\psi_n)$  are incompatible.  The characteristic
property  $\Pi(\psi_m)$ associated  with  state $\psi_m$  is a  binary
(yes/no) property  whose measurement asks  the question of  whether or
not $S$ is  in the state $\psi_m$.  The  two properties are compatible
by any one of the following criteria \cite{Fri10}: (a) in the sequence
of measurements $\Pi(\psi_m)\Pi(\psi_n)\Pi(\psi_m)$, the two instances
of   $\Pi(\psi_m)$  yield   the   same  outcome;   (b)  The   sequence
$\Pi(\psi_m)$ followed by  $\Pi(\psi_n)$ produces the same probability
distribution over  outcomes of $\Pi(\psi_n)$, as  a direct measurement
of $\Pi(\psi_n)$.

Thus, these  two measurements are incompatible  if acquiring knowledge
of   one  disturbs  the   other  and   diminishes  knowledge   of  it:
$H[\Pi(\psi_m)] \le H[\Pi(\psi_m)|  \Pi(\psi_n)]$, where $H(\cdot)$ is
Shannon  binary  entropy and  $H(\cdot|\cdot)$  is binary  conditional
entropy.  Two pure states  $\psi_m$ and $\psi_n$ are deemed compatible
if  and   only  if  the  associated   measurements  $\Pi(\psi_m)$  and
$\Pi(\psi_n)$   are  compatible.    The  incompatibility   implies  an
intrinsic randomness, i.e., one not having a deterministic explanation
within the theory \cite{AS3}.

We may extend this concept  of incompatibility of pure states to mixed
states  by associating  for a  state defined  by the  ensemble $\{p_j,
\psi_j\}$ the object $\sum_j  p_j \Pi(\psi_j)$.  In quantum mechanics,
the density operator $\rho$  naturally associates with a quantum state
in the  role of $\Pi(\{p_j,  \psi_j\})$, and the non-vanishing  of the
commutator of density operators is a ready witness of incompatibility.
An  interesting approach  to exposing  this non-classicality  by using
\textit{anti-commutators} is studied  in detail in Ref.  \cite{FMP+0}.
Here we consider a  particular quantification of nonclassicality based
on the commutator.

\section{Nonclassicality in quantum mechanics \label{sec:HS}}

The Hilbert-Schmidt  (HS) norm  of a (bounded,  square) operator  $A =
\{a_{jk}\}$ is given by
\begin{equation}
||A||^2_{HS}  \equiv \sum_{j,k}  |a_{jk}|^2  = \textrm{Tr}\left(A^\dag
A\right),
\label{eq:fro}
\end{equation}
where $a_{jk} \equiv \langle  j|A|k\rangle$ for any basis $|j\rangle$.
We assume that the dimension  is finite (though the arguments here can
be  generalized to infinite  dimension). Given  two states  $\rho$ and
$\sigma$, we propose the measure of their mutual incompatibility to be
twice the HS norm of the commutator:
\begin{equation}
\label{eq:Pi}
\Phi(\rho,\sigma) \equiv 2||[\rho, \sigma]||^2_{HS},
\end{equation}
where the pre-factor is required for normalization.  $\Phi$ so defined
is a convenient measure of  state incompatibility.  It is symmetric in
both  arguments,  and  its  interpretation  as  such  is  conceptually
transparent, while  it is computationally facile  (e.g., not involving
diagonalization).  Some of its  properties are studied below.  That it
brings  out the  intuitively expected  features of  nonclassicality is
shown later.

\begin{thm}
$0 \le \Phi(\rho,\sigma) \le 1$.
\end{thm}
\textbf{Proof.} Since  $\Phi(\rho,\sigma)$ is by  definition positive,
the  first inequality  in this  Theorem follows,  with  its saturation
precisely when $\rho$ and $\sigma$  are compatible. $\Phi$ is a convex
function in both  arguments, and attains its maximum  for pure states.
To see this,  let $\rho \equiv \sum_j p_j|\psi_j\rangle\langle\psi_j|$
and $\sigma \equiv  \sum_k q_k|\phi_k\rangle\langle\phi_k|$, where the
$|\psi_j\rangle$'s   and   $|\phi_k\rangle$'s   are  not   necessarily
orthogonal, and $\sum_j p_j  = \sum_k q_k=1$.  Let $\alpha_{jk} \equiv
\langle\psi_j|\phi_k\rangle$, $r_J \equiv  p_jq_k$, with $\sum_J r_J =
1$. We have:
\begin{widetext}
\begin{eqnarray}
\frac{1}{2}\Phi(\rho,\sigma) &=& \textrm{Tr}\left[
\left(\sum_{j,k} p_jq_k(\alpha_{jk}|\psi_j\rangle\langle\phi_k|
- \alpha_{jk}^\ast|\phi_j\rangle\langle\psi_k|)\right)^\dag
\left(\sum_{j^\prime,k^\prime} p_{j^\prime}q_{k^\prime}(\alpha_{j^\prime k^\prime}
|\psi_{j^\prime}\rangle\langle\phi_{k^\prime}|
- \alpha_{j^\prime k^\prime}^\ast|\phi_{j^\prime}
\rangle\langle\psi_{k^\prime}|)\right)
\right], \nonumber \\
&\equiv& \sum_{m,n}\langle m|
\left[\left(\sum_{J} r_JM_J^\dag\right)|n\rangle\langle n|
\left(\sum_{K} r_KM_K\right)\right]|m\rangle, \nonumber \\
&\equiv& \sum_{m,n}
\left(\sum_{J} r_JM_{J,mn}^\ast\right)
\left(\sum_{K} r_KM_{K,mn}\right) \nonumber \\
&\le&  \sum_J  r_J \sum_{m,n}  M_{J,mn}^\ast M_{J,mn} 
\equiv  \sum_J r_J
\sum_{m,n}\langle   m|   M_J^\dag|n\rangle\langle   n|   M_J|m\rangle,
\nonumber  \\
 &=&  \sum_{j,k}p_jq_k
\textrm{Tr}\left[      (\alpha_{jk}|\psi_j\rangle\langle\phi_k|      -
  \alpha_{jk}^\ast|\phi_j\rangle\langle\psi_k|)^\dag       (\alpha_{jk}
  |\psi_{j}\rangle\langle\phi_{k}|      -      \alpha_{jk}^\ast|\phi_j
  \rangle\langle\psi_{k}|) \right] \nonumber \\
&=& \frac{1}{2}\sum_{j,k}p_jq_k\Phi(\psi_j,\phi_k)
\label{eq:1} 
\end{eqnarray}
\end{widetext}
where we  used the fact that  $\sum_n |n\rangle\langle n|=\mathbb{I}$,
the        definition       $M_J        \equiv        M_{[jk]}       =
(\alpha_{jk}|\psi_j\rangle\langle\phi_k|                              -
\alpha_{jk}^\ast|\phi_j\rangle\langle\psi_k|)$  and  the convexity  of
the function $f(x) := |x|^2$.

Any two pure states $|\psi_1\rangle$ and $|\psi_2\rangle$ form a 2-dim
subspace and we can write without loss of generality $|\psi_2\rangle =
\cos\theta|\psi_1\rangle   +   \sin\theta|\psi_1^\perp\rangle$,  where
$\langle   \psi_1|\psi_1^\perp\rangle   =   0$.    Setting   $\rho   =
|\psi_1\rangle\langle        \psi_1|$       and        $\sigma       =
|\psi_2\rangle\langle\psi_2|$  in Eq.   (\ref{eq:Pi}),  and maximizing
over  $\theta$,  we  find   $\theta_{\rm  max}  =  \frac{\pi}{4}$  and
$\Phi_{\rm max} =1$.  \hfill $\blacksquare$

We note that  $\Phi$ does not attain its maximum  value for two states
selected from  a pair of  mutually unbiased bases (MUBs),  even though
MUBs  are  maximally non-commuting  in  the  sense  that the  entropic
uncertainty  relation  given  by  $H_P  +  H_Q  \ge  -2\log_2(|\langle
p|q\rangle|)$,  is the  most stringent  in  this case,  the rhs  being
$\log(d)$.   Here  $H_P$  ($H_Q$)  is  the  classical  binary  entropy
generated by measuring $P$  ($Q$), while $|\langle p|q\rangle|$ is the
largest   overlap   between   the   eigenvectors  of   $P$   and   $Q$
\cite{MU88}.  For  two  states  from  an MUB  pair,  without  loss  of
generality,  we   may  take  $\Phi\left(|0\rangle,  \frac{1}{\sqrt{d}}
\sum_{j=0}^{d-1}  |j\rangle\right) = \frac{4(d-1)}{d^2}$,  which falls
linearly with  dimension $d$.  This  of course happens because  as $d$
increases, these two vectors are increasingly mutually orthogonal, and
hence commuting.

Unlike  the anti-commutator,  which can  be measured  (possibly  by an
interative   procedure)    using   an   interferometer   \cite{FMP+0},
determining the value of the commutator experimentally requires a more
detailed  set-up, ideally  a  quantum tomography  of  the state.   For
sufficiently small  systems, this  is technologically feasible  at the
present  time.  E.g.,   Ref.   \cite{RLB+05}  reports  tomographically
determined quantum characteristics of a quantum walk on eight steps.

As an  application of $\Phi$,  we study below the  classicalization of
quantum walk,  the quantum  generalization of classical  random walks.
Because of quantum interference, the position probability distribution
of a QW deviates  from the classical linear-spreading Gaussian pattern
to a quadratic-spreading  twin-peaked pattern.  Adding noise gradually
imposes classical  behavior, returning it to  Gaussian behavior, which
has  been studied  by a  number  of authors  (Ref.  \cite{BSC+11}  and
references therein).   Because $\Phi$ is  a \textit{relative} measure,
we require  a set $\Sigma$ of  states of a system  $S$, with $|\Sigma|
\ge 2$,  to witness or quantify  the nonclassicality of  $S$.  Only if
$\Phi(\rho_j,\rho_k)$ vanishes (or, is sufficiently low) for all pairs
$\rho_j,\rho_k   \in  \Sigma$   ($j  \ne   k$)  can   $S$   be  called
classical. Otherwise, $S$ is nonclassical. Two strategies for choosing
$\Sigma$  and  quantifying nonclassicality  of  a noisy  time-evolving
system are considered below.

\section{Application to quantum walks \label{sec:QW}}

We model the linear discrete-time  (DT) quantum walker (QW) as a qubit
(coin) in Hilbert space $\mathcal{H}_C \equiv \textrm{span}(|0\rangle,
|1\rangle)$, that  can assume states in  position space $\mathcal{H}_P
\equiv \textrm{span}(|\psi_x\rangle)$,  where $x$ is  an integer.  The
linear walk  may be extended to  higher dimensions, as  well as assume
non-trivial  topologies,  such  as   a  cycle  (Ref.   \cite{CB0}  and
references therein).  The  state of the noisy QW  after $t$ time steps
is obtained iteratively according to:
\begin{eqnarray}
\rho(t) &=& \sum_{j_1,j_2,\cdots,j_t} A_{j_t}U_{j_t}\cdots A_{j_1}U_{j_1}\rho_0
U^\dag_{j_1}A^\dag_{j_1}\cdots U^\dag_{j_t}A^\dag_{j_t} \nonumber \\
    &\equiv &  \sum_{j_1,j_2,\cdots,j_t} \rho(t;j_1,j_2,\cdots,j_t),
\label{eq:QW}
\end{eqnarray}
through sequential applications of the coin-position unitary operation
$U_j$ and  the coin-specific noise  operation $\mathcal{E}$ determined
by  the Kraus  operators $A_j$.   Here  the initial  state is  $\rho_0
\equiv   |\Psi_0\rangle\langle\Psi_0|$,    where   $|\Psi_0\rangle   =
\frac{|0\rangle+ i |1\rangle}{\sqrt{2}}\otimes |{x=0}\rangle$. At each
time $t$, one applies the unitary $U \equiv W(C \otimes \mathbbm{1})$,
where   $C$  is   the  coin   operation   $\left(  \begin{array}{clcr}
  \cos(\alpha)  &   &  $~~~$  \sin(\alpha)  \\  \sin(\alpha)   &  &  -
  \cos(\alpha)
\end{array} \right)$ that rotates the state of the coin,
while $W$ shifts the position conditioned on coin state
\begin{equation}
W   \equiv  |0\rangle\langle  0|\otimes\sum_x|{x-1}\rangle\langle
x|  +  |1\rangle\langle  1|\otimes\sum_x|{x+1}\rangle\langle
x|.
\end{equation}
The position probability distribution at time $t$ is given by $ P(x) =
\textrm{Tr}_{PC}\left[\left(\Pi_x  \otimes  I_C\right)\rho(t)\right]$,
where $\Pi_x  \equiv |x\rangle\langle x|$  is the projector to  a wave
packet localized at position $x$.

For  the  noise  model,   we  choose  the  amplitude  damping  channel
\cite{NC00}, which describes a qubit interacting with a vacuum bath:
\begin{equation}
A_0 \equiv
\left( \begin{array}{cc}
\sqrt{1 - \mu} & 0 \\
0 & 1
\end{array}\right);~
A_1 \equiv
\left(
\begin{array}{cc}
0 & 0 \\
0 & \sqrt{\mu}
\end{array}\right),
\label{eq:AD}
\end{equation}
where $\mu  (\in [0,1])$ describes  the strength of the  noise.  Other
possible  models include  dephasing  noise or  more general  amplitude
damping  noise on  the coin,  such as  squeezed  generalized amplitude
damping \cite{SB08},  or dephasing in the position  degree of freedom.
The above simple noise model suffices for our present purpose.

\begin{figure}
\includegraphics[width=9.1cm]{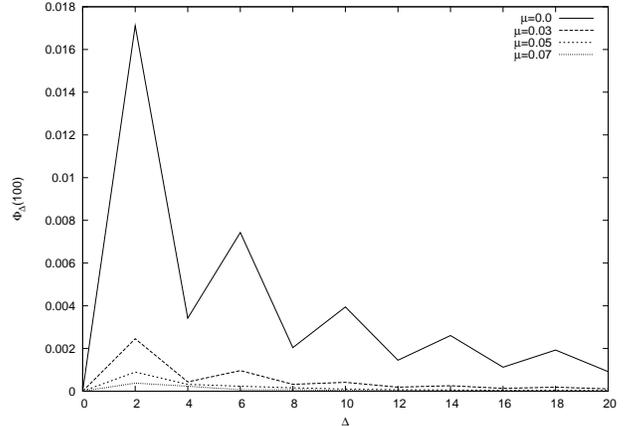}
\caption{Quantumness  $\Phi_\Delta(100)$ as  a function  of separation
  time  $\Delta$.  For the  system and  noise considered,  the optimal
  separation is seen to be $\Delta=2$.}
\label{fig:pfaz_tdiff}
\end{figure}

Two approaches may be considered to apply the $\Phi$ formalism. In one
case, $\Sigma(t)$  may be defined  as the ensemble of  states obtained
along different trajectories during the time interval $[0,t]$ starting
from $|\Psi_0\rangle$.   For the discrete-time evolution  given by Eq.
(\ref{eq:QW}),   one    considers   the   ensemble-dependent   average
quantumness
\begin{equation}
\Phi_{\rm          av}(t)         \equiv         \sum_{j_m,j^\prime_n}
\Phi\left[\rho(t;j_1,\cdots,j_t),\rho(t;j_1^\prime,\cdots,j_t^\prime)\right],
\end{equation}
where the unnormalized  density operators $\rho(t;j_1,\cdots,j_t)$ are
already factored by their statistical weight.

Another method,  which is used  here, would be to  consider $\Sigma(t)
\equiv   \{\rho(t),   \rho(t+\Delta)\}$,   where  $\rho(t)$   is   the
time-evolved mixed state density  operator of the system, and $\Delta$
is a time step that may be optimized to maximize $\Phi$. Thus:
\begin{equation}
\Phi_\Delta(t) \equiv \Phi\left(\rho(t),\rho(t-\Delta)\right),
\label{eq:Delta}
\end{equation}
where $\rho(t)$ is given by  Eq. (\ref{eq:QW}).  Keeping $t$ fixed, we
varied $\Delta$  to numerically determine the  $\Delta$ that maximizes
$\Phi_\Delta$.  We  find that $\Delta=2$  is optimal for  this system.
The data for $t=100$  is depicted in Figure \ref{fig:pfaz_tdiff}.  For
$\Delta=0$,  we  find  trivially  that $\Phi_\Delta=0$.   As  $\Delta$
increases, so  does $\Phi_\Delta$  as the state  of the QW  is rotated
away  from  $|\Psi_0\rangle$.  Eventually,  a  fall  with $\Delta$  is
expected  because the  dominant support  for two  QW states  will move
apart quadratically with  time, so that they will  nearly commute even
in the unitary case.

\begin{figure}
\includegraphics[width=9.1cm]{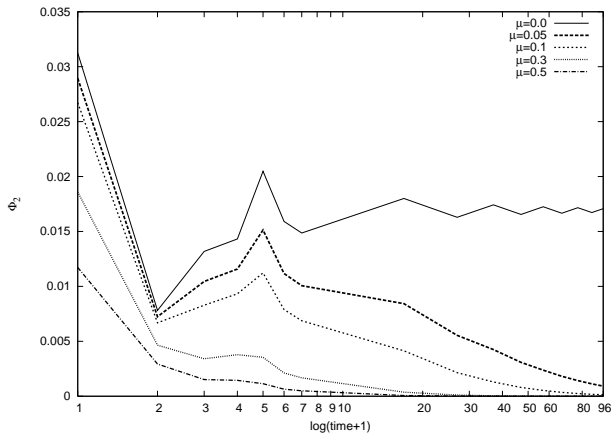}
\caption{Quantumness   $\Phi_2(t)$,   with   curves  parametrized   by
  different  values   of  the  amplitude   damping  channel  parameter
  $\mu$ in Eq. (\ref{eq:AD}).}
\label{fig:ampdamp_t}
\end{figure}
Figure  \ref{fig:ampdamp_t} depicts  $\Phi_\Delta(t)$  with $\Delta=2$
and  the  time being  varied  till  100 for  a  QW  described by  Eqs.
(\ref{eq:QW}) and (\ref{eq:AD}) for  various levels of noise $\mu$. We
find that at any given  time, quantumness is larger when the evolution
is unitary  (topmost plot), and  is successively smaller as  the noise
level increases.   For any fixed  noise level, quantumness is  seen to
reduce with time (the bottom  three plots), whereas it remains roughly
the same  when the walk  is unitary. These observations  give evidence
that $\Phi$ is  a reasonable measure of the  quantumness. 

\begin{figure}
\includegraphics[width=9.1cm]{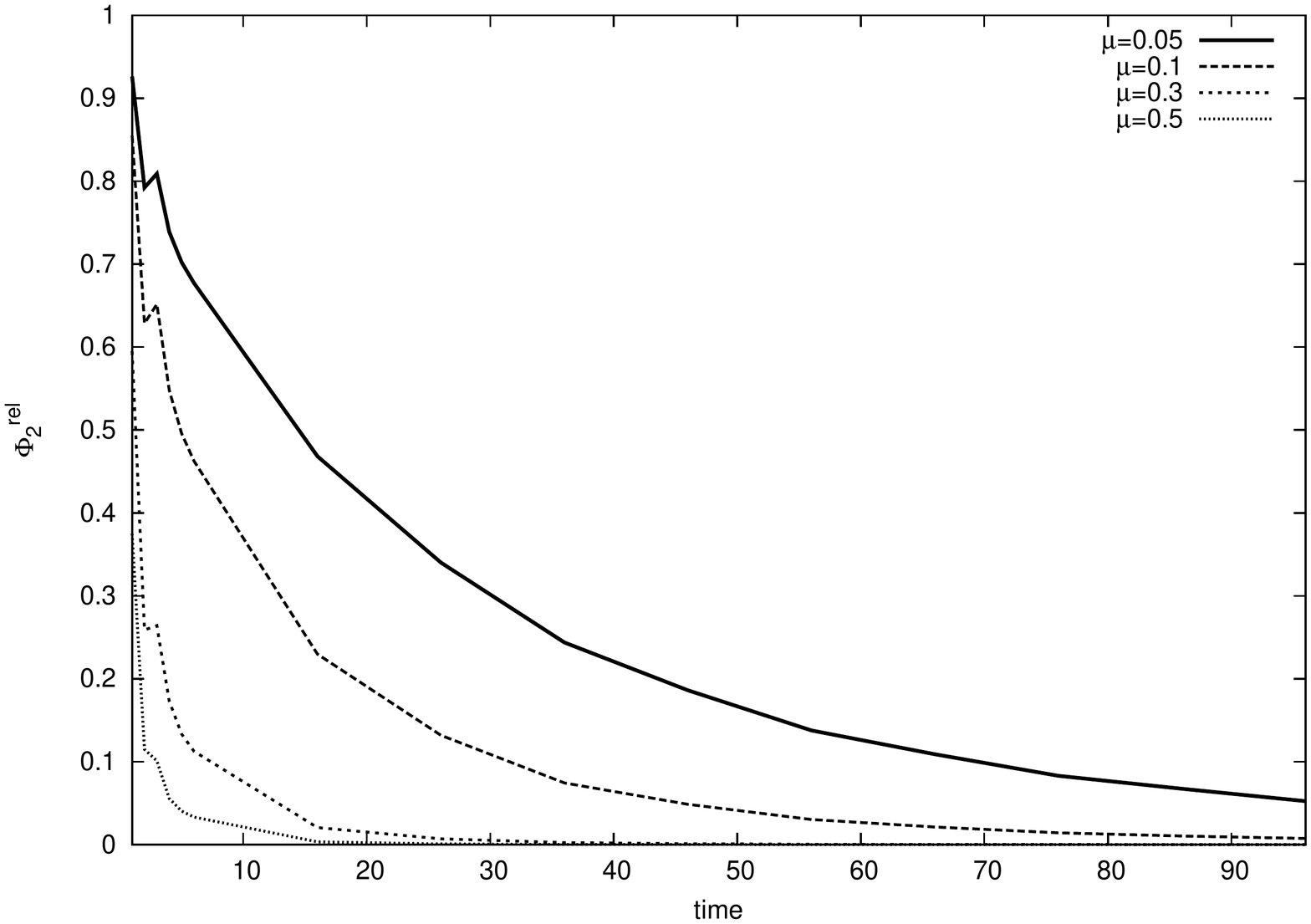}
\caption{The  relative quantumness  $\Phi^{\rm  rel}_2$, derived  from
  Fig.   \ref{fig:ampdamp_t}  by  normalizing  each  $\Phi_2(t)$  with
  respect to the noiseless value at the same time $t$.  The curves are
  parametrized by  different values  of the amplitude  damping channel
  parameter $\mu$ in  Eq.  (\ref{eq:AD}). Departure from 1  can now be
  directly interpreted as a sign of classicality.}
\label{fig:ampdamp_t_relative}
\end{figure}

In practice, $\Phi$ may be normalized by a suitable constant depending
on  the  system  $S$  at  hand.   As one  example,  we  plot  in  Fig.
(\ref{fig:ampdamp_t_relative}) a  time-normalized version of  the data
in Fig.  (\ref{fig:ampdamp_t}), where the noisy values of $\Phi_2$ are
divided  by the noiseless  value at  that time  $t$.  This  removes an
artefact of our method whereby a low $\Phi$ results not from noise but
from  the  fact  that the  compared  pair  of  states are  not  highly
non-commuting even in the unitary case.

\section{Concluding remarks \label{sec:C}}

The quantumness  of noisy quantum walks  has been studied  by means of
$\Phi$, applied  here to quantify the  non-commutativity of temporally
near-by states.  Applying this measure to  the case of DT linear QW to
which  an amplitude-damping  noise is  applied to  the coin  degree of
freedom, we show  that it brings out the  expected classicalization of
the walk,  thereby illustratating the quantitative  usefulness of this
intuitive   measure   of   quantumness.    It   can   be   implemented
experimentally  using quantum  state  tomography in  NMR systems,  and
potentially apply to study the quantumness of photosynthetic systems.

\acknowledgments

PI  and CGN  thank RRI  for  the support  through VSP  program and  PI
(Project  fellow) also  thanks  IISER Pune  through  the Project  code
no. 30111063.

\bibliography{vayu1}

\end{document}